\documentclass[%
 reprint,
 aps,
]{revtex4-2}

\usepackage{graphicx}
\usepackage{dcolumn}
\usepackage{bm}

\usepackage{comment}
\usepackage{amsmath}
\usepackage{amssymb}

\usepackage{xcolor}

\begin{document}

\preprint{APS/123-QED}

\title{Herd Behaviour in Public Goods Games}


\author{Mar\'ia Pereda}
\affiliation{Grupo de Investigación Ingenier\'ia de Organizaci\'on y Log\'istica (IOL), Departamento Ingenier\'ia de Organizaci\'on, Administraci\'on de empresas y Estad\'istica, Escuela T\'ecnica Superior de Ingenieros Industriales, Universidad Polit\'ecnica de Madrid, 28006 Madrid, Spain. \\Grupo Interdisciplinar de Sistemas Complejos (GISC), 28911 Legan\'es, Madrid, Spain}
 
\email{maria.pereda@upm.es}

\date{\today}

\begin{abstract}

The problem of free-riding arises when individuals benefit from a shared resource, service, or public good without contributing proportionately to its provision. This conduct often leads to a collective action problem, as individuals pursue personal gains while relying on the contributions of others. In this study, we present a Bayesian inference model to elucidate the behaviour of participants in a Public Goods Game, a conceptual framework that captures the essence of the free-riding problem. Here, individuals possess information on the distribution of group donations to the public good. Our model is grounded in the premise that individuals strive to harmonise their actions with the group's donation patterns. Our model is able to replicate behavioural patterns that resemble those observed in experiments with midsized groups (100 people), but fails to replicate those for larger scales (1000 people). Our results suggest that, in these scenarios, humans prefer imitation and convergence behaviours over profit optimisation. These insights contribute to understanding how cooperation is achieved through alignment with group behaviour.
\end{abstract}

\maketitle


\section{\label{sec:Intro} Introduction}


When exploring the complex dynamics of human cooperation and decision-making, the provision of public goods has attracted a great deal of attention in recent decades. At the core of this research lies the profound dilemma between individually rational decisions and the collective optimal outcome. The extensively researched paradigm for this issue is the Public Goods Game (PGG)~\cite{ledyard_1995}, where $N \geq 2$ agents can voluntarily contribute to a common pool. The pool is then multiplied or enhanced by a factor $1<r<N$, and the resulting value is shared among all participants, regardless of their individual contributions. 
The Nash Equilibrium in this game is characterised by investing zero amount in the public good, while the collective optimal outcome demands maximal contributions from all participants. Beyond its theoretical intrigue, this dilemma is of vital importance in addressing real-world challenges such as climate change mitigation, public infrastructure management, ecosystem protection, and sustainable exploitation, where solutions hinge on voluntary contributions from many. Despite game theory predictions (\cite{ModelsBasedPD, handbookEE, TheEconomyBook}, experimental findings, both in controlled laboratory settings and online, consistently show that participants contribute approximately half of their endowment in one-shot games; and, in iterated games, contributions decline but persist, with a general pattern of gradual decrease as players learn the consequences of their actions. The observed results are influenced by the information that participants receive about the actions of other players~\cite{zelmer_linear_2003, peng_punishment_2022, antonioni_know_2014, cuesta_reputation_2015, pereda_group_2019, pereda_large_2019}.

Much research has explored individual contributions in traditional Public Goods Games, where decisions are made after being informed of the last average contribution of the group members. Studies conducted by~\cite{fischbacher_are_2001,fischbacher_social_2010,li_conditional_2023} convincingly showcase the widespread occurrence of conditional cooperation in the Public Goods game. Using the strategy method, participants were categorised into two predominant groups: conditional cooperators, who are inclined to contribute more as others contribute, and free riders, who abstain from contributions regardless of others' behaviour. In conditions where individuals cannot base their decisions on others' behaviour, conditional cooperators, as demonstrated by ~\cite{fischbacher_are_2001,fischbacher_social_2010}, determine their contributions relying on diverse beliefs about the contributions of other group members. As far as we are aware, only one study~\cite{pereda_large_2019} has studied human decisions when presented with the distribution of decisions (not solely the average contribution of the group), and to date, no models have explained the behaviour observed in such a setup. Most of the modelling efforts have focused on understanding the group dynamics based on the average contribution to the group. Our model diverges from those standard cooperation models (see \cite{perc_evolutionary_2013, perc_statistical_2017} for two recent reviews on cooperation), from theoretical modelling approaches~\cite{gokhale_evolutionary_2010,han_equilibrium_2012,duong_analysis_2016,szolnoki_conditional_2012}, and models using the conditional cooperation mechanism \cite{fischbacher_are_2001, chaudhuri_sustaining_2011, Balaraju2020}. In our proposal, agents base their decisions on the distribution of decisions of the group, employing Bayesian inference to converge toward the group's patterns of contribution behaviour, as individuals possess pronounced inclinations to conform and mimic the group~\cite{boyd_richerson_2006,mesoudi_how_2009}.

Regarding the use of Bayesian inference to explain game-theoretical dilemmas, a large number of articles are devoted to the study of Bayesian games, that is, games with incomplete information~\cite{harsanyi_games_1967, aumann_repeated_1995, hal-02447604} and extensive research is focused on the field of decision-making under uncertainty. In~\cite{achtziger_fast_2014}, the authors built a model based on the psychological dual-process theories of intuition decision-making; combining two decision processes, a controlled one using Bayesian inference and an intuitive one based on reinforced learning. The model is used to draw conclusions about the frugality and quality of both types of decision-making processes. In the same uncertain context, the experimental results also showed that the most frequent decision rule used by people was Bayes's rule~\cite{ArePeopleBayesian}.

In this article, we present a model that attempts to explain the results observed in a non-traditional Public Goods Game experiment~\cite{pereda_large_2019} in which participants were given, each round, the distribution of donations from all members of the group, and given a window of 24 hours to make each round decision, allowing for well-thought decisions~\cite{achtziger_fast_2014}, which require time. The model we propose is grounded in the notion that individuals seek convergence towards the group's opinion~\cite{boyd_richerson_2006,mesoudi_how_2009}. Consequently, they adjust their information based on the collective decision-making process and decide using these updated beliefs. In essence, agents are not driven to optimise their individual or group gains throughout the game. Instead, they actively align themselves with the group's behaviour.

The paper is organised as follows. We present the definition of the model in Sec. II. The results obtained by simulation are analysed in Sec. III. Finally, we conclude with Sec. IV summarising the insights offered by our work.

\section{\label{sec:Models} MODEL DYNAMICS}

In this section, we present a model for the observed human behaviour under a Public Goods Game in which people are informed of the distribution of contributions of others. In the present model, agent donations are based on Bayesian inference using this information.

A crucial parameter in the linear Public Goods Games is the Marginal Per Capita Return (MPCR), which dictates the individual payoff for contributing an additional unit to the public good. A Prisoner's Dilemma situation arises when $1/N < MPCR = r/N < 1$, where $N$ is the number of decision-makers and $r$ is the multiplication factor. In cases where $MPCR < 1/N$, opting not to contribute becomes the payoff-dominant strategy. Conversely, if $MPCR > 1$, there is no dilemma and contributing becomes the payoff-dominant strategy. In our model, while we do not explicitly model or study the influence of $r$, we assume that the agents participate in a setup where $1/N < MPCR = r/N < 1$.

In our models, as in \cite{pereda_large_2019}, for each experimental round, $N$ agents receive 10 ECU (Experimental Conditioning Unit, the virtual currency inside the game) and decide how much to donate to the common pool: 0, 2, 4, 6, 8, 10 ECU (from zero to all the endowment). Therefore, the probability that an agent donates each of the six quantities follows a categorical distribution $\bm{X} \sim Categorical(\theta_{1}, \dots, \theta_{m})$, where $m$ represents the number of categories and $\theta_{i}$ the probability of choosing the category $i$. 


To make this decision, each Bayesian agent uses a prior distribution. In the four models, we use the Dirichlet distribution as prior, since it is a conjugate prior for the categorical distribution. Hence, $\bm{\theta} \sim Dir(\alpha_{1}, \dots, \alpha_{m})$ and 
\begin{equation}
P(\bm{\theta}) \propto \prod_{i=1}^{m}\theta_{i}^{\alpha_{i}-1} I_{\sum_{i=1}^{m} \theta_{i}=1, \theta_{i} \geq 0}
\label{eq:prior}
\end{equation}

where $I$ is the indicator function.

After each round, the agents observe the distribution of decisions of the group, i.e., the percentage of people that chose each category, and compute the posterior predictive distribution to make their new decision.

\vspace{4mm}

The model operates as follows:
\begin{enumerate}
    \item Initialise the prior $\bm{\theta} \sim Dir(\bm{\alpha})$.
    \item The agents make their first decision using the prior predictive distribution $\bm{Y}\sim Categorical(\theta_{1}, \dots, \theta_{m})$.
    \item From round 2 to the final round, the agents observe the decisions of the $N$ agents in the previous round (data), defining the likelihood for the data $P(Y|\bm{\theta}) \sim Categorical(\theta_{1}, \dots, \theta_{m})$. 
    \item The agents compute the posterior distribution, which is proportional to the likelihood $\times$ the prior
    \begin{eqnarray}
    P(\bm{\theta}|Y) \propto P(Y|\bm{\theta}) \times P(\bm{\theta})= \nonumber \\ \prod_{i=1}^{m}\theta_{i}^{c_{i}+\alpha_{i}-1} \propto Dir(\bm{\theta}|\bm{c}+\bm{\alpha})
    \label{eq:posterior}
    \end{eqnarray}
    where the vector $\bm{c}$ summarises the number of decisions in each category present in the data.
    \item The agents compute the posterior predictive distribution 
    \begin{eqnarray}
    P(X'|Y)=\int P(X'|\bm{\theta},Y) \times P(\bm{\theta}|Y)d\theta = \nonumber \\
    \int P(X'|\bm{\theta}) \times P(\bm{\theta}|Y)d\theta = \mathbb{E}(\bm{\theta}_{X'}|Y)= \nonumber \\
    (c_x + \alpha_x) \mathbin{/} (n + \alpha_0)
    \label{eq:posteriorPred}
    \end{eqnarray}
    where $c_x$ counts the number of decisions in category x, and $\alpha_{0}=\sum_{i=1}^{m} \alpha_{i}$.
    \item The agents make their next decision using the posterior predictive distribution. Then we repeat steps 3 to 6 until the last round is reached.
\end{enumerate}

\vspace{4mm}

The model is implemented in four versions, with different parametrizations. Models (A) and (B) differ on the initial conditions: model (A) uses a non-informative prior for the agents, and model (B) uses a data-based prior. The last two versions are motivated by the present of ``special individuals'' in real experimental settings, i.e., there is always a percentage of people who do not respond to external feedback or information but always behave as full cooperator or free rider (they donate all or nothing, respectively). We include this more realistic composition of agents to the model. Consequently, these third and fourth models derive from models (A) and (B) including non-Bayesian agents that behave as full cooperators or free riders. Then, the third model (C) includes non-Bayesian agents and uses the data-based prior from (B), and the fourth one (D) includes non-Bayesian agents and the non-informative prior from (A).

The model versions have been implemented using the \emph{R} language, and the code can be downloaded from github \url{https://github.com/mpereda/Herding_in_PGG}. 

In the following, we present the four models.

\subsection{\label{sec:Model1}Bayesian agents and non-informative prior}

In this initial model, we assume that the agents start their participation in the game with no initial knowledge of the behaviour of people in this type of game. Therefore, we parameterise the Dirichlet distribution for the prior to be the least informative prior possible, with an invariant distribution, so we use the Jeffreys prior. The Jeffreys prior for the categorical distribution is a Dirichlet distribution with parameters $\alpha_{i}=1/2$ for $i=1,...,m$.
Therefore, the prior for the agents follows Eq.~(\ref{eq:prior}) with $\alpha_{i}=1/2$.

\subsection{\label{sec:Model2}Model with Bayesian agents and data-based prior}

In this second model, we assume the individuals possess knowledge about the behaviour of people in this kind of dilemmas, and so can estimate a prior distribution of donations to use in their first decision. This would be in line with the observed first donations and donations in one-shot games, which happen to be approximately half of the endowment for a great variety of games (see~\cite{Sanchez_2018} for a review on the topic). In the model, we use empirical data from~\cite{pereda_large_2019} which can be found in the Zenodo repository that accompanies the referred paper. The prior distribution is then a Dirichlet distribution with parameters $\alpha_{1}=0.1$, $\alpha_{2}=0.2$, $\alpha_{3}=0.32$, $\alpha_{4}=0.2$, $\alpha_{5}=0.07$, and $\alpha_{6}=0.11$ (see details in the code that accompanies this manuscript).

\subsection{\label{sec:Model3}Model with Bayesian agents, data-based prior, and non-Bayesian agents}

In this third version, we incorporate non-Bayesian agents: full cooperators and free riders. We estimate the percentage of full cooperators and free riders from the data in ~\cite{pereda_large_2019}. To do so, we define \emph{full cooperator} as someone who donated the whole endowment in more than 80\% rounds, and, analogously, we define the \emph{free rider} as those people who donated zero in more than 80\% of the rounds. Then, we compute the average number of full cooperators and free riders in the experimental data, resulting in a 5\% of full cooperators and 6\% of free riders (for details, see the code accompanying this work).

\subsection{\label{sec:Model4}Model with Bayesian agents, non-informative prior, and non-Bayesian agents}

This last model incorporates the non-informative prior from model (A) and includes the same percentage of non-Bayesian agents in model (C).



\section{\label{sec:Results} Results}

This section presents the simulation results of 100 replications for each model, with populations of N=100 and N=1000 agents. 


For each model, we perform cluster analysis on the simulation results (agent behaviours) to explore whether the models can generate different types of aggregated behavioural patterns, as seen in the experimental data (e.g., highly polarised behaviours or mean-centred behaviours). Specifically, we perform hierarchical agglomerative clustering with Euclidean distance and average linkage (using the R library \emph{hclust}). Each data instance is an 84-dimensional vector that summarises each simulation run, where position $d_{i*(j-6)}$ represents the percentage of agents making the $i-th$ decision at round $j$.

In the supplementary material, we provide, for each model, a histogram of histograms (aggregating data from multiple model replications) using scatterplot-type graphs with jitter. For each cluster in the cluster analyses, we provide examples of instances inside the clusters.

\begin{figure}[htp]
\includegraphics[width=80mm]{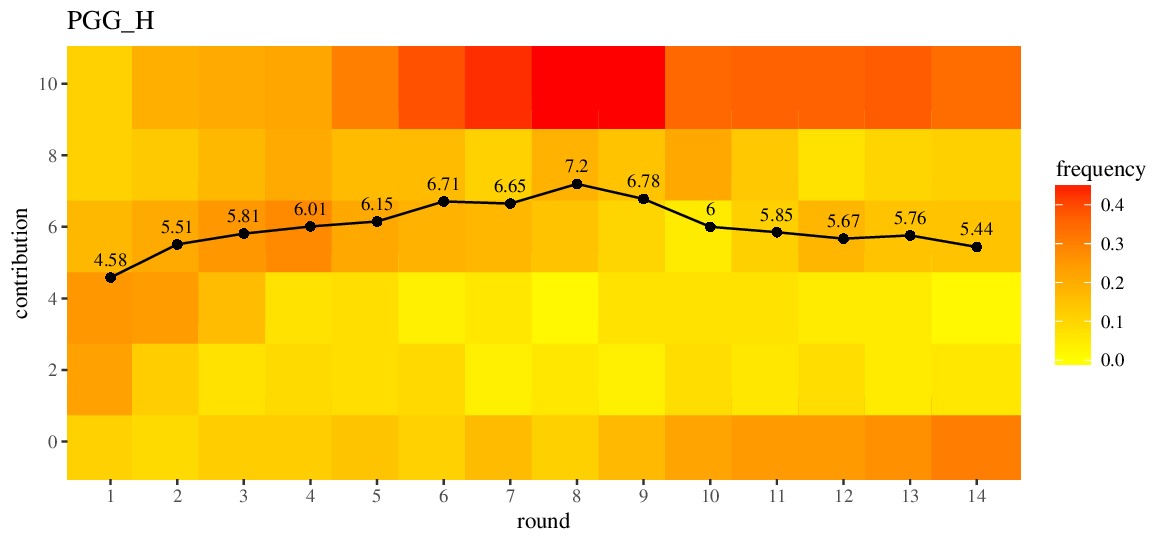}
\caption{\label{fig:Heatmap1} Heatmap of the distribution of decisions in a 100-people PGG experiment with feedback on the distribution of donations (PGG\_H), reproduced from \cite{pereda_large_2019} with permission.}
\end{figure}

\begin{figure}[htp]
\includegraphics[width=80mm]{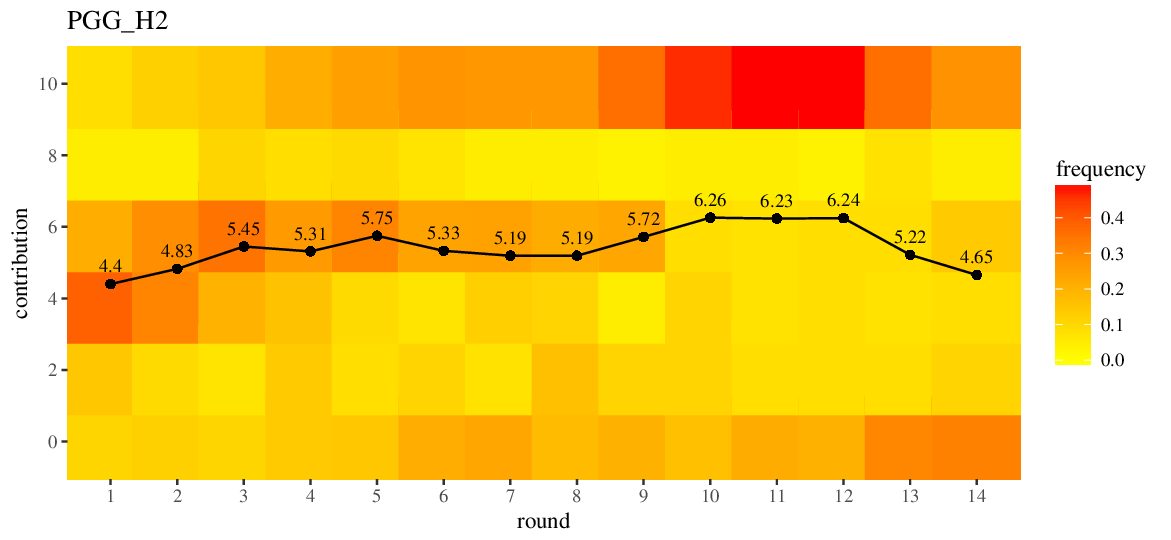}
\caption{\label{fig:Heatmap2} Heatmap of the distribution of decisions in a second 100-people PGG experiment with feedback on the distribution of donations (PGG\_H2), reproduced from \cite{pereda_large_2019} with permission.}
\end{figure}

The experimental results we use for comparison, from \cite{pereda_large_2019}, are shown in Fig.\ref{fig:Heatmap1} and Fig.\ref{fig:Heatmap2}, which have been reproduced with the permission of the authors.

\subsection{\label{sec:RModel1}Results for the basic model with Bayesian agents and noninformative prior}

After conducting 100 simulations with N=100 agents and N=1000 agents, we proceeded to perform a cluster analysis on them. An overview of all simulations is presented in Supplementary Figures S1 and S2. 

Cluster analysis unveils the presence of six distinct behaviour clusters, labelled as clusters 1 through 6. Their representativeness, quantified by the percentage of replications belonging to each cluster, is as follows: 31\% for cluster 1, 32\% for cluster 2, 11\% for cluster 3, 21\% for cluster 4, 3\% for cluster 5, and 2\% for cluster 6. See Supplementary Figures S3 to S8 for simulation examples of each cluster. 

\begin{figure}[htp]
\includegraphics[width=80mm]{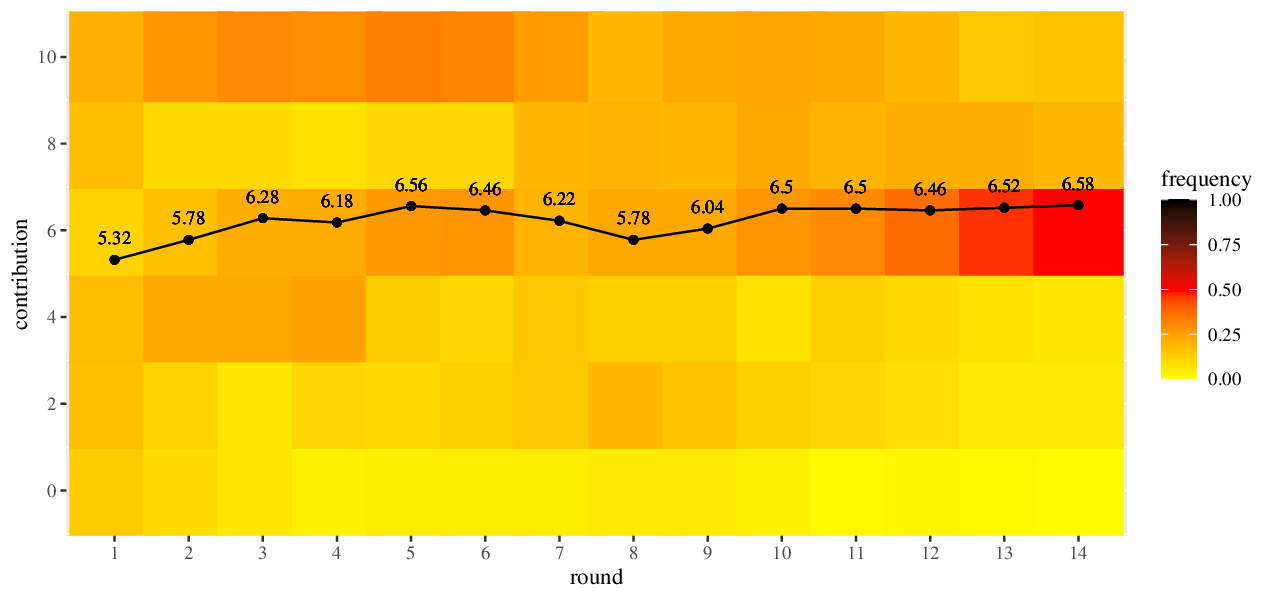}
\caption{\label{fig:Model1_100_cluster1_rep9} Heatmap of the distribution of donations of a simulation of cluster type 1 behaviour of the basic model (A) with N=100. The average donation is shown as dots joined with lines, in black.}
\end{figure}

\begin{figure}[htp]
\includegraphics[width=80mm]{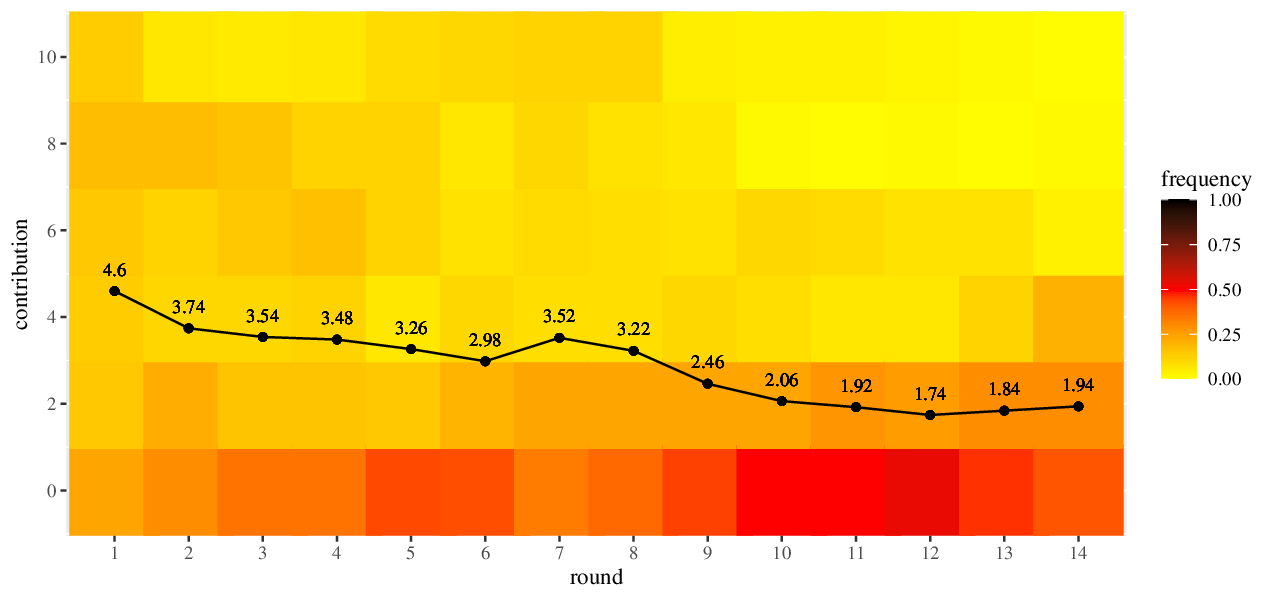}
\caption{\label{fig:Model1_100_cluster2_rep12} Heatmap of the distribution of donations of a simulation of cluster type 2 behaviour of the basic model (A) with N=100. The average donation is shown as dots joined with lines, in black.}
\end{figure}

Cluster 1 comprises simulations in which the most common cooperative behaviour is centred around the average contribution of the group, along with simulations in which the majority of contributions are above the mean endowment (5 points); see Fig.~\ref{fig:Model1_100_cluster1_rep9} as a reference. In Cluster 2, simulations are notable for the presence of a high percentage of agents contributing low values of the endowment, with a non-negligible number of simulations featuring a very high percentage of agents who do not contribute at all (Fig.~\ref{fig:Model1_100_cluster2_rep12}). Cluster 4 groups simulations in which the predominant contribution is 20\% of the endowment (2 points) (see Supplementary Figure S6). The remaining clusters encapsulate less frequent behaviours (see Supplementary Figures S5, S7 and S8).

In simulations with N=1000 agents, a predominant cluster emerges, which includes 98\% of the simulations, in which individual behaviours have homogenized (Fig.~\ref{fig:Model1_1000_cluster1_rep38}). See Supplementary Figures S2 and S9-S10 for simulation examples from each cluster.

\begin{figure}[htp]
\includegraphics[width=80mm]{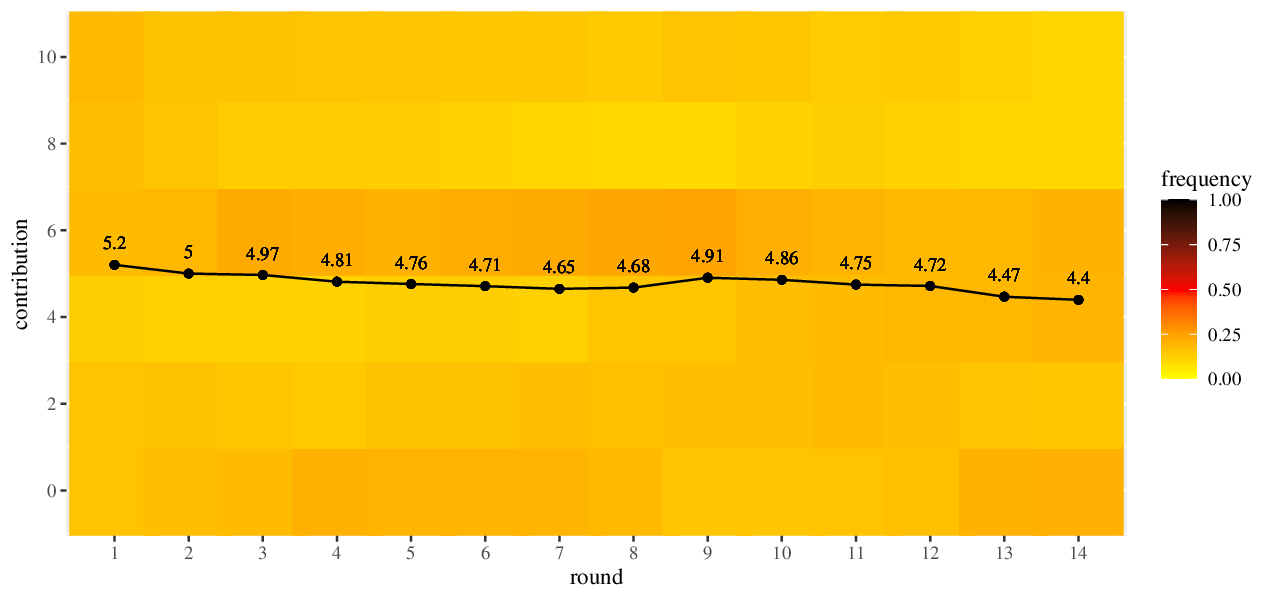}
\caption{\label{fig:Model1_1000_cluster1_rep38} Heatmap of the distribution of donations of a simulation of the principal cluster for N=1000 of the basic model (A). The average donation is shown as dots joined with lines, in black.}
\end{figure}

It can be observed that a population size that is not sufficiently large prevents Bayesian agents from homogenising their behaviour, as they would do if they had sufficient information with a larger sample size, as in the case of N=1000.

\subsection{\label{sec:RModel2}Results for the model with Bayesian agents and data-based prior}

An encompassing display of all simulations can be found in Supplementary Figures S11 (N=100) and S12 (N=1000).

\begin{figure}[htp]
\includegraphics[width=80mm]{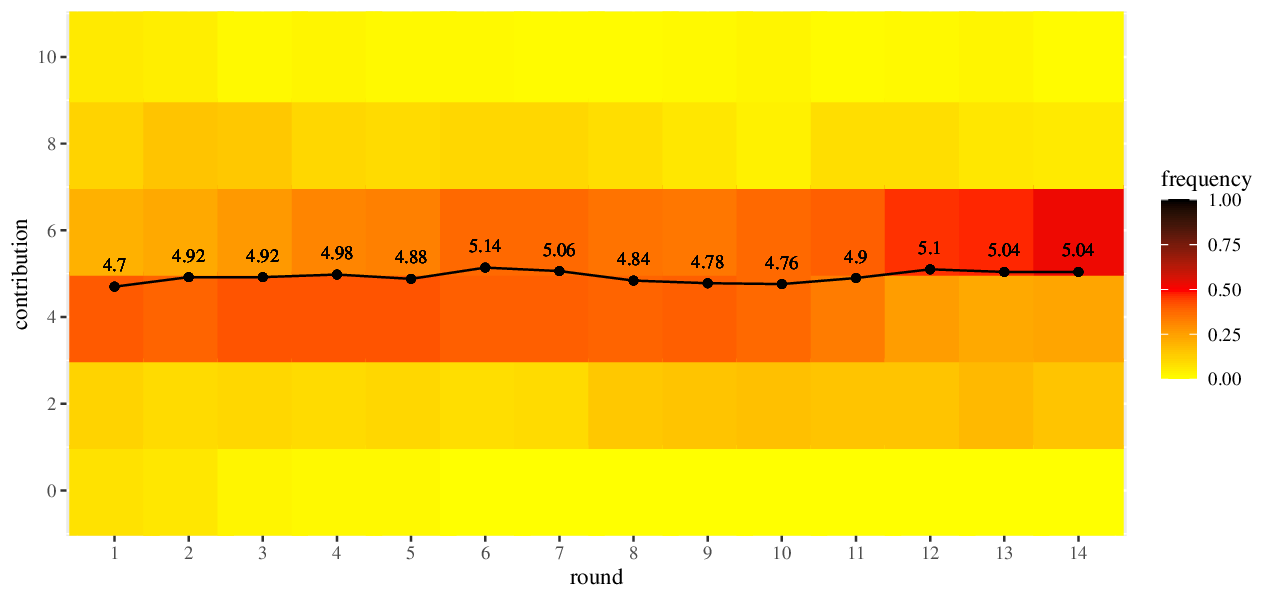}
\caption{\label{fig:Model2_100_cluster1_rep1} Heatmap of the distribution of donations of a simulation of cluster type 1 behaviour of the model with Bayesian agents and data-based prior (B) with N=100. The average donation is shown as dots joined with lines, in black.}
\end{figure}

\begin{figure}[htp]
\includegraphics[width=80mm]{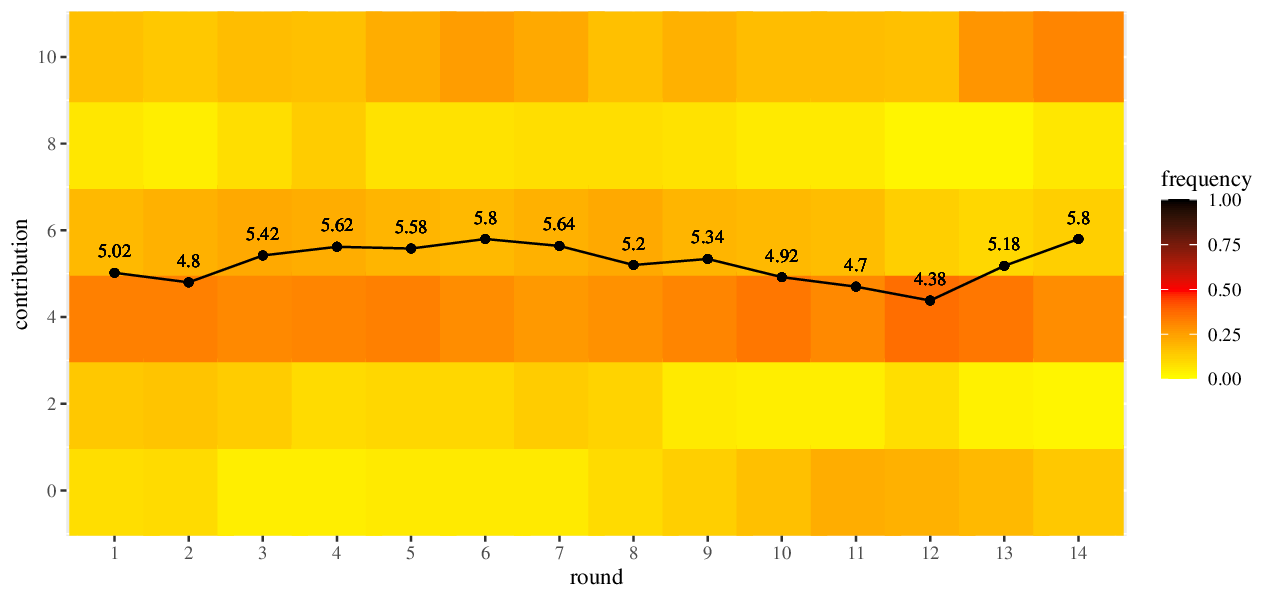}
\caption{\label{fig:Model2_100_cluster2_rep77} Heatmap of the distribution of donations of a simulation of cluster type 2 behaviour of the model with Bayesian agents and data-based prior (B) with N=100. The average donation is shown as dots joined with lines, in black.}
\end{figure}

Cluster analysis reveals that simulations involving N=100 agents exhibit distinctive behavioural patterns, with six distinct categories emerging with the following percentages of representation: 49, 36, 1, 4, 4 and 6\%. The primary cluster, depicted in Supplementary Figure S13, constitutes 49\% of the behaviours and is characterised by cooperation values that converge around half of the endowment (see, for example, Fig.~\ref{fig:Model2_100_cluster1_rep1}). A secondary group, illustrated in S14, represents 36\% of the cases and displays patterns centred around half of the endowment but with higher dispersion, with some examples showing high percentages of agents donating the full endowment (Fig.~\ref{fig:Model2_100_cluster2_rep77}). The fourth cluster includes replications with the majority of agents contributing 40\% of the endowment (Figure S15). The fifth group shows pronounced polarisation towards significantly lower levels of cooperation (Figure S16). Lastly, the sixth cluster demonstrates polarisation towards higher cooperation values.

In the simulations involving N=1000 agents, a predominant cluster, as depicted in Supplementary Figure S18 and Figure \ref{fig:Model2_1000_cluster2_rep36.eps}, comes to the forefront, encompassing a remarkable 97\% of the simulations. Within this cluster, individual behaviours have significantly converged and homogenised around the distribution used as a prior.

\begin{figure}[htp]
\includegraphics[width=80mm]{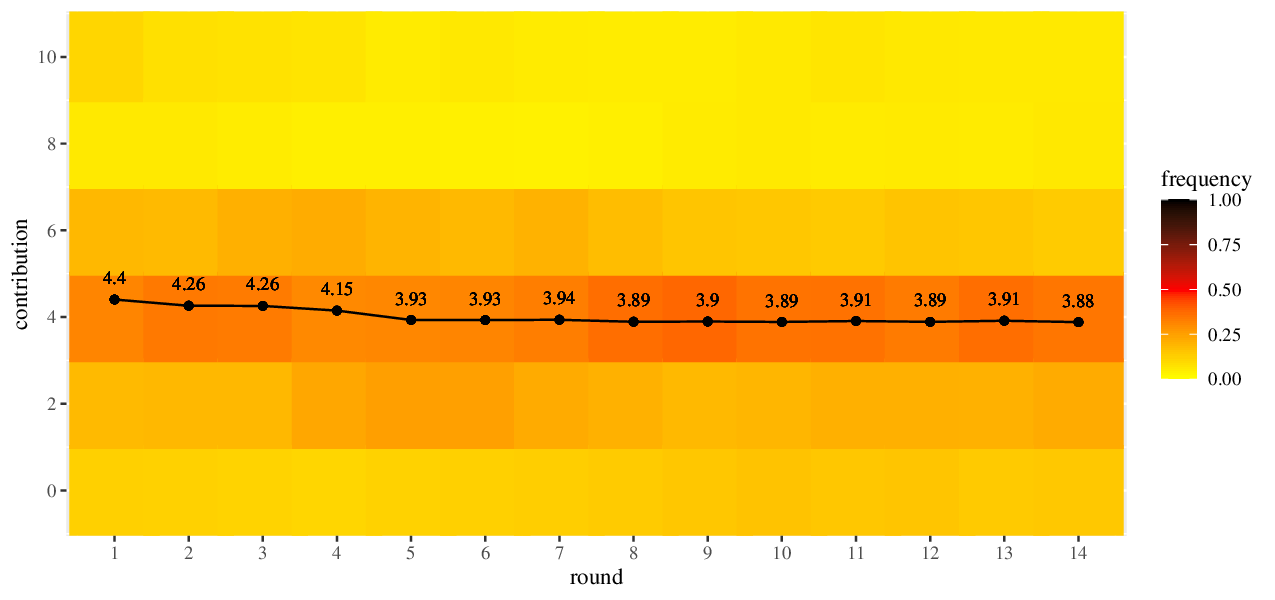}
\caption{\label{fig:Model2_1000_cluster2_rep36.eps} Heatmap of the distribution of donations of a simulation of cluster type 1 of the model with Bayesian agents and data-based prior (B) with N=1000. The average donation is shown as dots joined with lines, in black.}
\end{figure}

\subsection{\label{sec:RModel3}Results for the model with Bayesian agents, data-based prior, and non-Bayesian agents}

\begin{figure}[htp]
\includegraphics[width=80mm]{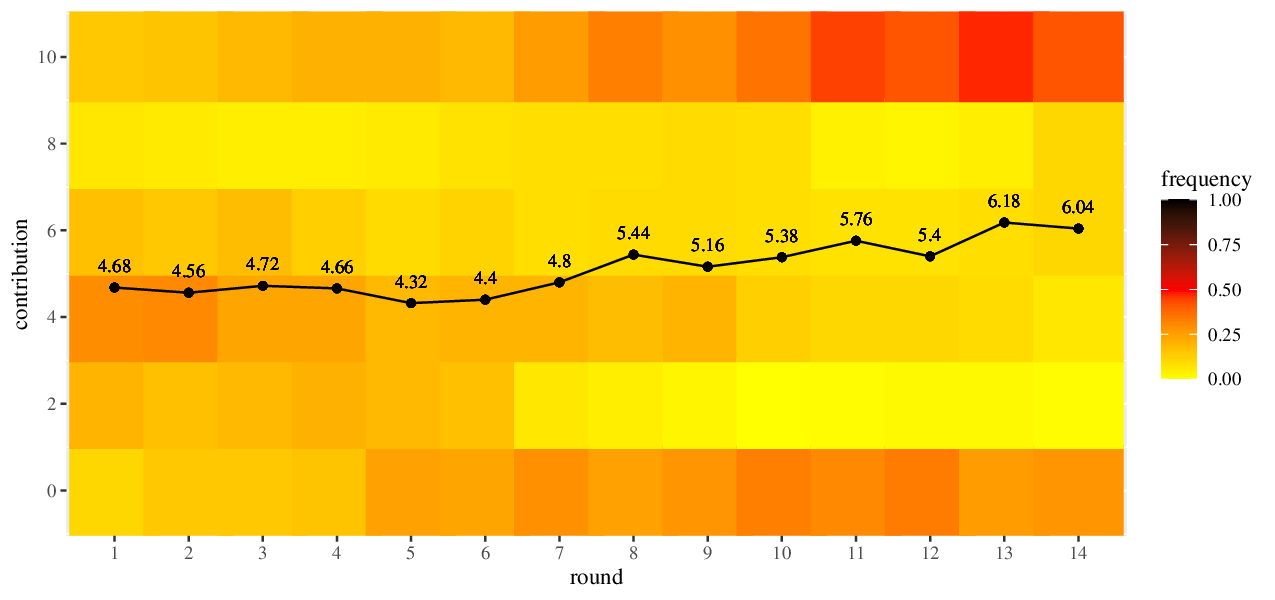}
\caption{\label{fig:Model3_100_cluster1_rep82} Heatmap of the distribution of donations of a simulation of cluster type 1 behaviour of the model with Bayesian and non-Bayesian agents and data-based prior (C) with N=100. The average donation is shown as dots joined with lines, in black.}
\end{figure}

The results of the cluster analysis with N=100 return 4 clusters (86\%, 6\%, 2\%, and 6\%, respectively), shown in the Supplementary Figures S21 to S24. The behaviours are grouped into a main cluster that contains 86\% of the simulations. These show agent behaviour that becomes more polarised as the rounds of the game progress towards highly polarised values (most agents donate all or nothing at the end of the game); see Fig.~\ref{fig:Model3_100_cluster1_rep82} as an example. Cluster 2 groups simulations with ultracooperative behaviour, i.e. those with an extremely high percentage of agents donating all the endowment (Fig. S22). The remaining clusters, which are less representative, show polarised simulations but to a lesser extent than the previous ones.

\begin{figure}[htp]
\includegraphics[width=80mm]{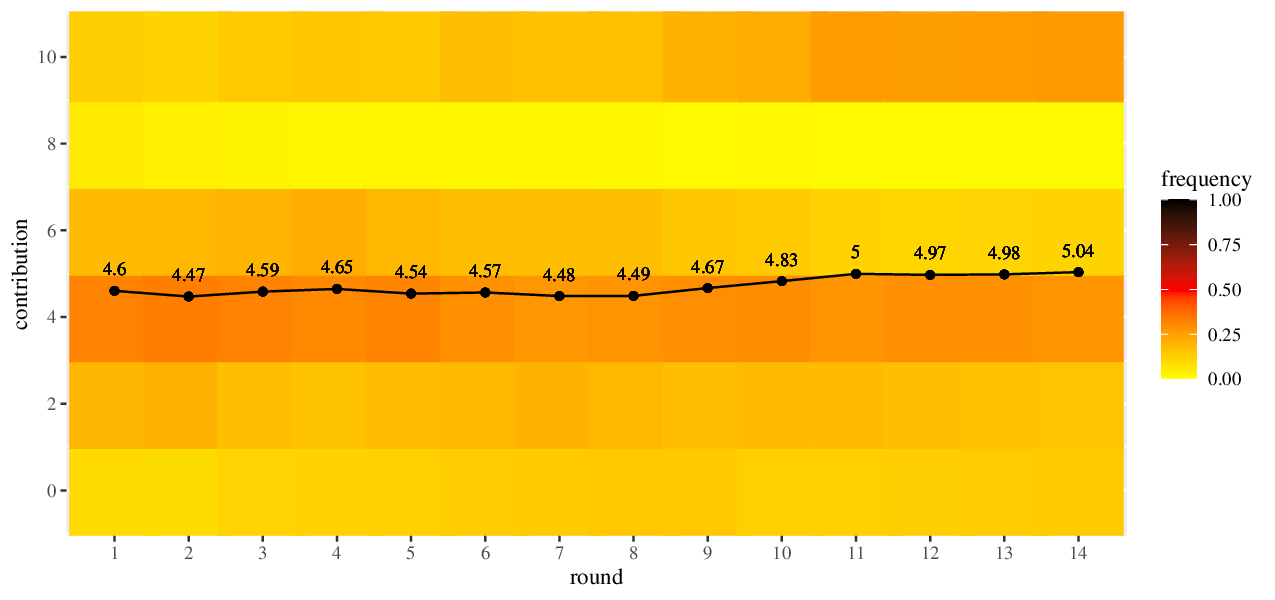}
\caption{\label{fig:Model3_1000_cluster1_rep71} Heatmap of the distribution of donations of a simulation of cluster type 1 behaviour of the model with Bayesian and non-Bayesian agents and data-based prior (C) with N=1000. The average donation is shown as dots joined with lines, in black.}
\end{figure}

The outcomes of cluster analysis for N=1000 reveal four clusters, which make up 81\%, 17\%, 1\%, and 1\% of the total, respectively. See Supplementary Figures S25 to S28 for more details. A cluster appears that groups most of the simulations 81\%, in which the agents' behaviours are grouped around average values of the endowment, as reflected in the prior used in the first round of the game. Non-negligible percentages of agents donating all or none of the endowment are observed, but in much lower proportions than in the case with N=100 agents. For reference, see the example in Fig.~\ref{fig:Model3_1000_cluster1_rep71}. The second cluster shows very similar behaviour, but with a higher percentage of donations concentrated in 40\% of the endowment (Supplementary Figure S26).

An overview of all simulations is provided in Supplementary Figures S19 and S20.

\subsection{\label{sec:RModel4}Results for the model with Bayesian agents, non-informative prior, and non-Bayesian agents}

Supplementary Figures S29 and S30 offer a comprehensive presentation of all simulations.

\begin{figure}[htp]
\includegraphics[width=80mm]{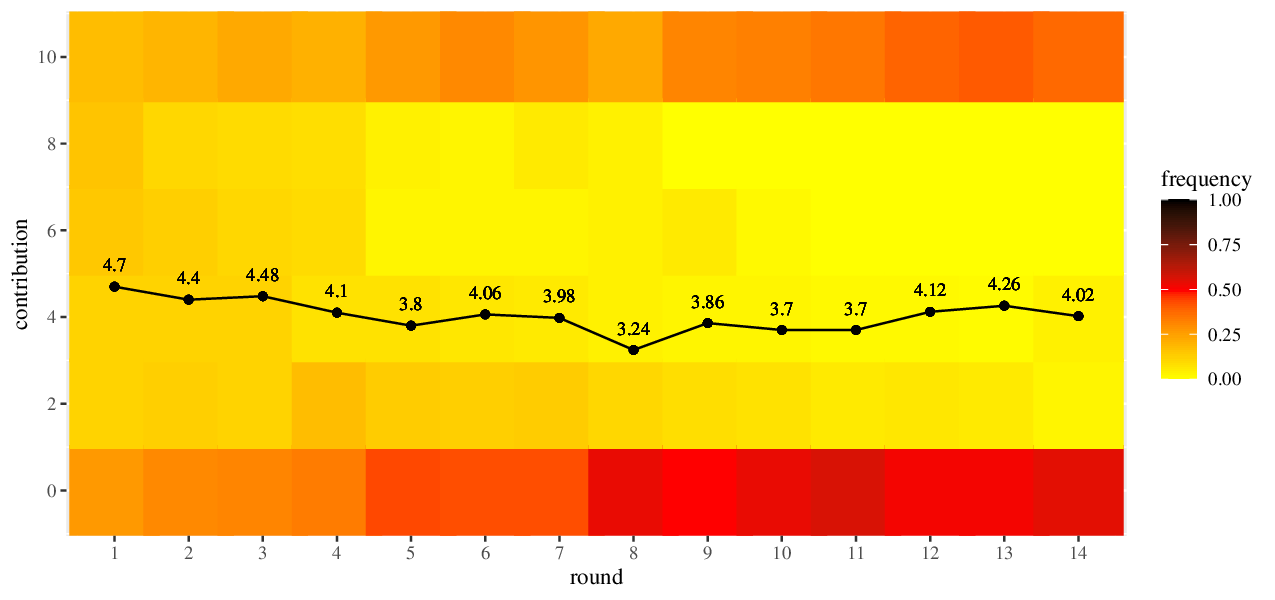}
\caption{\label{fig:Model4_100_cluster1_rep61} Heatmap of the distribution of donations of a simulation of cluster type 1 behaviour of the model with Bayesian and non-Bayesian agents and non-informative prior (D) with N=100. The average donation is shown as dots joined with lines, in black.}
\end{figure}

The results of the cluster analysis with N=100 indicate the presence of six clusters, representing 82\%, 5\%, 9\%, 1\%, 1\%, and 2\%, respectively. Supplementary Figures S31 to S36 provide a visual representation of these clusters. The majority cluster, with 82\% of the simulations, shows ultrapolarised behaviour, due to the presence of \emph{free riders} and \emph{full cooperators} (see Fig~\ref{fig:Model4_100_cluster1_rep61} and S31). With a non-informative prior, the dynamics of the agents are attracted by agents who do not change their opinion. The remaining clusters have simulations that are similar to cluster 1, with some particularities. They can be seen in Supplementary Figures S32 to S36.

\begin{figure}[htp]
\includegraphics[width=80mm]{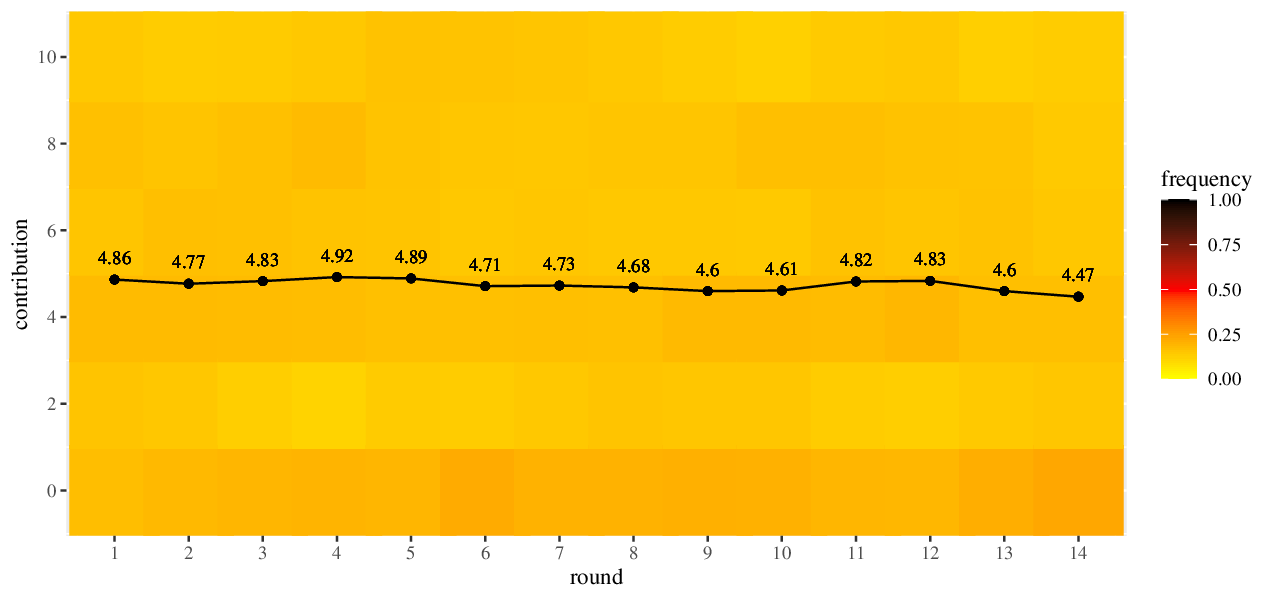}
\caption{\label{fig:Model4_1000_cluster} Heatmap of the distribution of donations of a simulation of cluster type 1 behaviour of the model with Bayesian and non-Bayesian agents and non-informative prior (D) with N=1000. The average donation is shown as dots joined with lines, in black.}
\end{figure}

The cluster analysis carried out with N=1000 delineates clusters with very similar instances, so we can consider that all behaviours have homogenised. Figure \ref{fig:Model4_1000_cluster} and Supplementary Figure S37 visually illustrate the characteristics of this unique group.

\section{\label{sec:Conclusions} Discussion and conclusions}

In this study, we have introduced a model to elucidate human behaviour in the Public Goods Game when individuals possess more information than just the average contribution of the group, incorporating details on the distribution of donations from all participants in the previous round. The model assumption is that individuals in this scenario are not seeking to optimise their gains, but rather to converge and imitate \emph{the group's} behaviour. The model is implemented in four different versions (A to D), each with distinct parameterizations. The first model, model (A), is the simplest, assuming that individuals have no knowledge of how others behave in this game. Model (B) refines this assumption, incorporating some form of knowledge, drawing insights from real experimental data~\cite{pereda_large_2019}. The last two models extend these initial models by introducing individuals who are not influenced by the group but exhibit independent behaviours such as \emph{free riders} or \emph{full cooperators}, as observed in experiments.

Overall, our results demonstrate scale effects on behaviours. Various patterns emerge for the distribution of donations in mid-sized groups (N=100), some resembling those observed in real human data. However, these patterns tend to disappear on larger scales (N = 1000), where a predominant and more homogeneous behavioural pattern is observed across replications.

Specifically, Model (C) with 100 agents exhibits behavioural patterns compatible with those experimentally observed in groups of 100 individuals (see Figures \ref{fig:Heatmap1} and \ref{fig:Heatmap1} vs \ref{fig:Model3_100_cluster1_rep82} and Supplementary Figure S21). Here, we observe how agent behaviour and human behaviour in the initial rounds stem from a prior assumption and subsequently evolve, polarising towards situations with a high percentage of full cooperators and free riders.

These findings could be attributed not only to the effect of Bayesian agents but also to a small-scale effect. In fact, we note that for larger group sizes, as seen in the results with 1000 agents presented throughout the study, the agent behaviours become much more homogeneous across all models. This homogenisation of behaviours in large populations directly results from employing larger samples for Bayesian inference. Since individuals update their priors throughout the game's evolution using observational data from the previous round, a larger population implies a larger data sample to estimate the true distribution of group decisions, and this larger sample expedites convergence to a steady state. This steady-state result to be the initial condition (plus the percentage of defectors and free riders when present).

In fact, on larger scales, such as with N = 1000 individuals, the model outcomes do not align with the experimental findings. This discrepancy arises because the experimental results from the sole replication involving 1000 participants resemble those observed with only 100 individuals. Consequently, drawing definitive conclusions and establishing causation proves challenging given the limited empirical evidence currently available. One potential explanation for this discrepancy may lie in individuals' inability to effectively compute and estimate priors and posteriors in groups larger than those that human's brains can handle, as suggested by Dunbar's number~\cite{DUNBAR1992469}, which is 150. The Dunbar's number refers to a theoretical cognitive limit to the number of people with whom one can maintain stable social relationships. Beyond this limit, it becomes increasingly difficult for individuals to maintain meaningful relationships and track social dynamics effectively.

Our results suggest that humans, in Public Goods Game scenarios, might prefer imitation and convergence behaviours over profit optimisation, providing a plausible explanation for the surprising results, high levels of average cooperation and polarisation, in Public Goods Dilemmas of this kind (individuals with more information). Indeed, our results align with previous findings on group conformity. Herding behaviour has been shown to be influenced by several factors, namely payoff externalities, correlated effects, information externalities, and social preferences. Payoff externalities manifest when an individual's actions impact the payoffs of others, leading to a collective choice and equilibrium~\cite{Choi97, Diamond83,Scharfstein90}. Similarly, correlated effects arise when agents adopt similar behaviours due to shared external influences~\cite{Manski2000}. In our work, agents are exposed to information externalities, where previous works showed that individuals base their decisions on information inferred from observing the choices of their predecessors, leading them to disregard their private information. Consequently, their decisions become uninformative for subsequent players, initiating a cascade effect~\cite{Banerjee92, Bikhchandani92, Welch92, CORAZZINI200774, Goeree2015}. In the psychology literature, individuals are also assumed to have a preference for conformity~\cite{Jones94, Bernheim94}.

Finally, adhering to Occam's Razor, our model, while limited, offer generic insights into how cooperation is achieved through the convergence to group behaviour. A potential extension of the model would be introducing noise or \emph{trembling hand} in decisions, as found in~\cite{ArePeopleBayesian}, where `people are prone to make errors and random decisions'. Simulations with this added noise would study which level of noise is able to destroy cooperation and how cooperation (distribution and average) and group conformity are affected by different levels of \emph{trembling hand}. This avenue presents an intriguing prospect for further exploration.

\begin{acknowledgments}
The genesis of this article emerged from informal discussions with Jos\'e A. Cuesta and Ignacio Tamarit during the GISC workshop in the pre-pandemic era. Thanks to both for sparking up the initial idea.
\end{acknowledgments}

\bibliography{apssamp}

\begin{thebibliography}{41}%
\makeatletter
\providecommand \@ifxundefined [1]{%
 \@ifx{#1\undefined}
}%
\providecommand \@ifnum [1]{%
 \ifnum #1\expandafter \@firstoftwo
 \else \expandafter \@secondoftwo
 \fi
}%
\providecommand \@ifx [1]{%
 \ifx #1\expandafter \@firstoftwo
 \else \expandafter \@secondoftwo
 \fi
}%
\providecommand \natexlab [1]{#1}%
\providecommand \enquote  [1]{``#1''}%
\providecommand \bibnamefont  [1]{#1}%
\providecommand \bibfnamefont [1]{#1}%
\providecommand \citenamefont [1]{#1}%
\providecommand \href@noop [0]{\@secondoftwo}%
\providecommand \href [0]{\begingroup \@sanitize@url \@href}%
\providecommand \@href[1]{\@@startlink{#1}\@@href}%
\providecommand \@@href[1]{\endgroup#1\@@endlink}%
\providecommand \@sanitize@url [0]{\catcode `\\12\catcode `\$12\catcode `\&12\catcode `\#12\catcode `\^12\catcode `\_12\catcode `\%12\relax}%
\providecommand \@@startlink[1]{}%
\providecommand \@@endlink[0]{}%
\providecommand \url  [0]{\begingroup\@sanitize@url \@url }%
\providecommand \@url [1]{\endgroup\@href {#1}{\urlprefix }}%
\providecommand \urlprefix  [0]{URL }%
\providecommand \Eprint [0]{\href }%
\providecommand \doibase [0]{https://doi.org/}%
\providecommand \selectlanguage [0]{\@gobble}%
\providecommand \bibinfo  [0]{\@secondoftwo}%
\providecommand \bibfield  [0]{\@secondoftwo}%
\providecommand \translation [1]{[#1]}%
\providecommand \BibitemOpen [0]{}%
\providecommand \bibitemStop [0]{}%
\providecommand \bibitemNoStop [0]{.\EOS\space}%
\providecommand \EOS [0]{\spacefactor3000\relax}%
\providecommand \BibitemShut  [1]{\csname bibitem#1\endcsname}%
\let\auto@bib@innerbib\@empty
\bibitem [{\citenamefont {Ledyard}(1995)}]{ledyard_1995}%
  \BibitemOpen
  \bibfield  {author} {\bibinfo {author} {\bibfnamefont {J.~O.}\ \bibnamefont {Ledyard}},\ }\bibinfo {title} {Public goods: A survey of experimental research},\ in\ \href@noop {} {\emph {\bibinfo {booktitle} {The Handbook of Experimental Economics}}}\ (\bibinfo  {publisher} {Princeton University Press},\ \bibinfo {year} {1995})\ pp.\ \bibinfo {pages} {111--194},\ \bibinfo {note} {funding by JPL}\BibitemShut {NoStop}%
\bibitem [{\citenamefont {Doebeli}\ and\ \citenamefont {Hauert}(2005)}]{ModelsBasedPD}%
  \BibitemOpen
  \bibfield  {author} {\bibinfo {author} {\bibfnamefont {M.}~\bibnamefont {Doebeli}}\ and\ \bibinfo {author} {\bibfnamefont {C.}~\bibnamefont {Hauert}},\ }\bibfield  {title} {\bibinfo {title} {Models of cooperation based on the prisoner's dilemma and the snowdrift game},\ }\href {https://doi.org/https://doi.org/10.1111/j.1461-0248.2005.00773.x} {\bibfield  {journal} {\bibinfo  {journal} {Ecology Letters}\ }\textbf {\bibinfo {volume} {8}},\ \bibinfo {pages} {748} (\bibinfo {year} {2005})}\BibitemShut {NoStop}%
\bibitem [{\citenamefont {Kagel}\ and\ \citenamefont {Roth}(2016)}]{handbookEE}%
  \BibitemOpen
  \bibfield  {author} {\bibinfo {author} {\bibfnamefont {J.}~\bibnamefont {Kagel}}\ and\ \bibinfo {author} {\bibfnamefont {A.}~\bibnamefont {Roth}},\ }\href@noop {} {\emph {\bibinfo {title} {The Handbook of Experimental Economics, Volume 2}}},\ \bibinfo {edition} {1st}\ ed.,\ Vol.~\bibinfo {volume} {2}\ (\bibinfo  {publisher} {Princeton University Press},\ \bibinfo {year} {2016})\BibitemShut {NoStop}%
\bibitem [{\citenamefont {Bowles}\ \emph {et~al.}(2023)\citenamefont {Bowles}, \citenamefont {Carlin},\ and\ \citenamefont {Stevens}}]{TheEconomyBook}%
  \BibitemOpen
  \bibfield  {author} {\bibinfo {author} {\bibfnamefont {S.}~\bibnamefont {Bowles}}, \bibinfo {author} {\bibfnamefont {W.}~\bibnamefont {Carlin}},\ and\ \bibinfo {author} {\bibfnamefont {M.}~\bibnamefont {Stevens}},\ }\href {https://www.core-econ.org/the-economy/microeconomics/04-strategic-interactions-06-public-good-games.html} {\emph {\bibinfo {title} {4.6 Public good games and cooperation. Unit 4 in The CORE Econ Team 2023 The Economy 2.0 (online)}}}\ (\bibinfo  {publisher} {the CORE project},\ \bibinfo {year} {2023})\BibitemShut {NoStop}%
\bibitem [{\citenamefont {Zelmer}(2003)}]{zelmer_linear_2003}%
  \BibitemOpen
  \bibfield  {author} {\bibinfo {author} {\bibfnamefont {J.}~\bibnamefont {Zelmer}},\ }\bibfield  {title} {\bibinfo {title} {Linear {Public} {Goods} {Experiments}: {A} {Meta}-{Analysis}},\ }\href {https://doi.org/10.1023/A:1026277420119} {\bibfield  {journal} {\bibinfo  {journal} {Experimental Economics}\ }\textbf {\bibinfo {volume} {6}},\ \bibinfo {pages} {299} (\bibinfo {year} {2003})}\BibitemShut {NoStop}%
\bibitem [{\citenamefont {Peng}(2022)}]{peng_punishment_2022}%
  \BibitemOpen
  \bibfield  {author} {\bibinfo {author} {\bibfnamefont {H.-C.}\ \bibnamefont {Peng}},\ }\bibfield  {title} {\bibinfo {title} {Punishment mechanisms and cooperation in public goods games: {Experimental} evidence},\ }\href {https://doi.org/10.1111/apce.12343} {\bibfield  {journal} {\bibinfo  {journal} {Annals of Public and Cooperative Economics}\ }\textbf {\bibinfo {volume} {93}},\ \bibinfo {pages} {533} (\bibinfo {year} {2022})},\ \bibinfo {note} {publisher: John Wiley \& Sons, Ltd}\BibitemShut {NoStop}%
\bibitem [{\citenamefont {Antonioni}\ \emph {et~al.}(2014)\citenamefont {Antonioni}, \citenamefont {Cacault}, \citenamefont {Lalive},\ and\ \citenamefont {Tomassini}}]{antonioni_know_2014}%
  \BibitemOpen
  \bibfield  {author} {\bibinfo {author} {\bibfnamefont {A.}~\bibnamefont {Antonioni}}, \bibinfo {author} {\bibfnamefont {M.~P.}\ \bibnamefont {Cacault}}, \bibinfo {author} {\bibfnamefont {R.}~\bibnamefont {Lalive}},\ and\ \bibinfo {author} {\bibfnamefont {M.}~\bibnamefont {Tomassini}},\ }\bibfield  {title} {\bibinfo {title} {Know {Thy} {Neighbor}: {Costly} {Information} {Can} {Hurt} {Cooperation} in {Dynamic} {Networks}},\ }\href {https://doi.org/10.1371/journal.pone.0110788} {\bibfield  {journal} {\bibinfo  {journal} {PLOS ONE}\ }\textbf {\bibinfo {volume} {9}},\ \bibinfo {pages} {e110788} (\bibinfo {year} {2014})},\ \bibinfo {note} {publisher: Public Library of Science}\BibitemShut {NoStop}%
\bibitem [{\citenamefont {Cuesta}\ \emph {et~al.}(2015)\citenamefont {Cuesta}, \citenamefont {Gracia-Lázaro}, \citenamefont {Ferrer}, \citenamefont {Moreno},\ and\ \citenamefont {Sánchez}}]{cuesta_reputation_2015}%
  \BibitemOpen
  \bibfield  {author} {\bibinfo {author} {\bibfnamefont {J.~A.}\ \bibnamefont {Cuesta}}, \bibinfo {author} {\bibfnamefont {C.}~\bibnamefont {Gracia-Lázaro}}, \bibinfo {author} {\bibfnamefont {A.}~\bibnamefont {Ferrer}}, \bibinfo {author} {\bibfnamefont {Y.}~\bibnamefont {Moreno}},\ and\ \bibinfo {author} {\bibfnamefont {A.}~\bibnamefont {Sánchez}},\ }\bibfield  {title} {\bibinfo {title} {Reputation drives cooperative behaviour and network formation in human groups},\ }\href {https://doi.org/10.1038/srep07843} {\bibfield  {journal} {\bibinfo  {journal} {Scientific Reports}\ }\textbf {\bibinfo {volume} {5}},\ \bibinfo {pages} {7843} (\bibinfo {year} {2015})}\BibitemShut {NoStop}%
\bibitem [{\citenamefont {Pereda}\ \emph {et~al.}(2019{\natexlab{a}})\citenamefont {Pereda}, \citenamefont {Capraro},\ and\ \citenamefont {Sánchez}}]{pereda_group_2019}%
  \BibitemOpen
  \bibfield  {author} {\bibinfo {author} {\bibfnamefont {M.}~\bibnamefont {Pereda}}, \bibinfo {author} {\bibfnamefont {V.}~\bibnamefont {Capraro}},\ and\ \bibinfo {author} {\bibfnamefont {A.}~\bibnamefont {Sánchez}},\ }\bibfield  {title} {\bibinfo {title} {Group size effects and critical mass in public goods games},\ }\href {https://doi.org/10.1038/s41598-019-41988-3} {\bibfield  {journal} {\bibinfo  {journal} {Scientific Reports}\ }\textbf {\bibinfo {volume} {9}},\ \bibinfo {pages} {5503} (\bibinfo {year} {2019}{\natexlab{a}})}\BibitemShut {NoStop}%
\bibitem [{\citenamefont {Pereda}\ \emph {et~al.}(2019{\natexlab{b}})\citenamefont {Pereda}, \citenamefont {Tamarit}, \citenamefont {Antonioni}, \citenamefont {Cuesta}, \citenamefont {Hernández},\ and\ \citenamefont {Sánchez}}]{pereda_large_2019}%
  \BibitemOpen
  \bibfield  {author} {\bibinfo {author} {\bibfnamefont {M.}~\bibnamefont {Pereda}}, \bibinfo {author} {\bibfnamefont {I.}~\bibnamefont {Tamarit}}, \bibinfo {author} {\bibfnamefont {A.}~\bibnamefont {Antonioni}}, \bibinfo {author} {\bibfnamefont {J.~A.}\ \bibnamefont {Cuesta}}, \bibinfo {author} {\bibfnamefont {P.}~\bibnamefont {Hernández}},\ and\ \bibinfo {author} {\bibfnamefont {A.}~\bibnamefont {Sánchez}},\ }\bibfield  {title} {\bibinfo {title} {Large scale and information effects on cooperation in public good games},\ }\href {https://doi.org/10.1038/s41598-019-50964-w} {\bibfield  {journal} {\bibinfo  {journal} {Scientific Reports}\ }\textbf {\bibinfo {volume} {9}},\ \bibinfo {pages} {15023} (\bibinfo {year} {2019}{\natexlab{b}})}\BibitemShut {NoStop}%
\bibitem [{\citenamefont {Fischbacher}\ \emph {et~al.}(2001)\citenamefont {Fischbacher}, \citenamefont {Gächter},\ and\ \citenamefont {Fehr}}]{fischbacher_are_2001}%
  \BibitemOpen
  \bibfield  {author} {\bibinfo {author} {\bibfnamefont {U.}~\bibnamefont {Fischbacher}}, \bibinfo {author} {\bibfnamefont {S.}~\bibnamefont {Gächter}},\ and\ \bibinfo {author} {\bibfnamefont {E.}~\bibnamefont {Fehr}},\ }\bibfield  {title} {\bibinfo {title} {Are people conditionally cooperative? {Evidence} from a public goods experiment},\ }\href {https://doi.org/10.1016/S0165-1765(01)00394-9} {\bibfield  {journal} {\bibinfo  {journal} {Economics Letters}\ }\textbf {\bibinfo {volume} {71}},\ \bibinfo {pages} {397} (\bibinfo {year} {2001})}\BibitemShut {NoStop}%
\bibitem [{\citenamefont {Fischbacher}\ and\ \citenamefont {Gächter}(2010)}]{fischbacher_social_2010}%
  \BibitemOpen
  \bibfield  {author} {\bibinfo {author} {\bibfnamefont {U.}~\bibnamefont {Fischbacher}}\ and\ \bibinfo {author} {\bibfnamefont {S.}~\bibnamefont {Gächter}},\ }\bibfield  {title} {\bibinfo {title} {Social {Preferences}, {Beliefs}, and the {Dynamics} of {Free} {Riding} in {Public} {Goods} {Experiments}},\ }\href {https://doi.org/10.1257/aer.100.1.541} {\bibfield  {journal} {\bibinfo  {journal} {American Economic Review}\ }\textbf {\bibinfo {volume} {100}},\ \bibinfo {pages} {541} (\bibinfo {year} {2010})}\BibitemShut {NoStop}%
\bibitem [{\citenamefont {Li}\ and\ \citenamefont {Noussair}(2023)}]{li_conditional_2023}%
  \BibitemOpen
  \bibfield  {author} {\bibinfo {author} {\bibfnamefont {T.}~\bibnamefont {Li}}\ and\ \bibinfo {author} {\bibfnamefont {C.~N.}\ \bibnamefont {Noussair}},\ }\bibfield  {title} {\bibinfo {title} {Conditional cooperation and group size: experimental evidence from a public good game},\ }\bibfield  {journal} {\bibinfo  {journal} {Journal of the Economic Science Association}\ }\href {https://doi.org/10.1007/s40881-023-00152-4} {10.1007/s40881-023-00152-4} (\bibinfo {year} {2023})\BibitemShut {NoStop}%
\bibitem [{\citenamefont {Perc}\ \emph {et~al.}(2013)\citenamefont {Perc}, \citenamefont {Gómez-Gardeñes}, \citenamefont {Szolnoki}, \citenamefont {Floría},\ and\ \citenamefont {Moreno}}]{perc_evolutionary_2013}%
  \BibitemOpen
  \bibfield  {author} {\bibinfo {author} {\bibfnamefont {M.}~\bibnamefont {Perc}}, \bibinfo {author} {\bibfnamefont {J.}~\bibnamefont {Gómez-Gardeñes}}, \bibinfo {author} {\bibfnamefont {A.}~\bibnamefont {Szolnoki}}, \bibinfo {author} {\bibfnamefont {L.~M.}\ \bibnamefont {Floría}},\ and\ \bibinfo {author} {\bibfnamefont {Y.}~\bibnamefont {Moreno}},\ }\bibfield  {title} {\bibinfo {title} {Evolutionary dynamics of group interactions on structured populations: a review},\ }\href {https://doi.org/10.1098/rsif.2012.0997} {\bibfield  {journal} {\bibinfo  {journal} {Journal of The Royal Society Interface}\ }\textbf {\bibinfo {volume} {10}},\ \bibinfo {pages} {20120997} (\bibinfo {year} {2013})}\BibitemShut {NoStop}%
\bibitem [{\citenamefont {Perc}\ \emph {et~al.}(2017)\citenamefont {Perc}, \citenamefont {Jordan}, \citenamefont {Rand}, \citenamefont {Wang}, \citenamefont {Boccaletti},\ and\ \citenamefont {Szolnoki}}]{perc_statistical_2017}%
  \BibitemOpen
  \bibfield  {author} {\bibinfo {author} {\bibfnamefont {M.}~\bibnamefont {Perc}}, \bibinfo {author} {\bibfnamefont {J.~J.}\ \bibnamefont {Jordan}}, \bibinfo {author} {\bibfnamefont {D.~G.}\ \bibnamefont {Rand}}, \bibinfo {author} {\bibfnamefont {Z.}~\bibnamefont {Wang}}, \bibinfo {author} {\bibfnamefont {S.}~\bibnamefont {Boccaletti}},\ and\ \bibinfo {author} {\bibfnamefont {A.}~\bibnamefont {Szolnoki}},\ }\bibfield  {title} {\bibinfo {title} {Statistical physics of human cooperation},\ }\href {https://doi.org/10.1016/j.physrep.2017.05.004} {\bibfield  {journal} {\bibinfo  {journal} {Statistical physics of human cooperation}\ }\textbf {\bibinfo {volume} {687}},\ \bibinfo {pages} {1} (\bibinfo {year} {2017})}\BibitemShut {NoStop}%
\bibitem [{\citenamefont {Gokhale}\ and\ \citenamefont {Traulsen}(2010)}]{gokhale_evolutionary_2010}%
  \BibitemOpen
  \bibfield  {author} {\bibinfo {author} {\bibfnamefont {C.~S.}\ \bibnamefont {Gokhale}}\ and\ \bibinfo {author} {\bibfnamefont {A.}~\bibnamefont {Traulsen}},\ }\bibfield  {title} {\bibinfo {title} {Evolutionary games in the multiverse},\ }\href {https://doi.org/10.1073/pnas.0912214107} {\bibfield  {journal} {\bibinfo  {journal} {Proceedings of the National Academy of Sciences}\ }\textbf {\bibinfo {volume} {107}},\ \bibinfo {pages} {5500} (\bibinfo {year} {2010})},\ \bibinfo {note} {publisher: Proceedings of the National Academy of Sciences}\BibitemShut {NoStop}%
\bibitem [{\citenamefont {Han}\ \emph {et~al.}(2012)\citenamefont {Han}, \citenamefont {Traulsen},\ and\ \citenamefont {Gokhale}}]{han_equilibrium_2012}%
  \BibitemOpen
  \bibfield  {author} {\bibinfo {author} {\bibfnamefont {T.~A.}\ \bibnamefont {Han}}, \bibinfo {author} {\bibfnamefont {A.}~\bibnamefont {Traulsen}},\ and\ \bibinfo {author} {\bibfnamefont {C.~S.}\ \bibnamefont {Gokhale}},\ }\bibfield  {title} {\bibinfo {title} {On equilibrium properties of evolutionary multi-player games with random payoff matrices},\ }\href {https://doi.org/10.1016/j.tpb.2012.02.004} {\bibfield  {journal} {\bibinfo  {journal} {Theoretical Population Biology}\ }\textbf {\bibinfo {volume} {81}},\ \bibinfo {pages} {264} (\bibinfo {year} {2012})}\BibitemShut {NoStop}%
\bibitem [{\citenamefont {Duong}\ and\ \citenamefont {Han}(2016)}]{duong_analysis_2016}%
  \BibitemOpen
  \bibfield  {author} {\bibinfo {author} {\bibfnamefont {M.~H.}\ \bibnamefont {Duong}}\ and\ \bibinfo {author} {\bibfnamefont {T.~A.}\ \bibnamefont {Han}},\ }\bibfield  {title} {\bibinfo {title} {Analysis of the expected density of internal equilibria in random evolutionary multi-player multi-strategy games},\ }\href {https://doi.org/10.1007/s00285-016-1010-8} {\bibfield  {journal} {\bibinfo  {journal} {Journal of Mathematical Biology}\ }\textbf {\bibinfo {volume} {73}},\ \bibinfo {pages} {1727} (\bibinfo {year} {2016})}\BibitemShut {NoStop}%
\bibitem [{\citenamefont {Szolnoki}\ and\ \citenamefont {Perc}(2012)}]{szolnoki_conditional_2012}%
  \BibitemOpen
  \bibfield  {author} {\bibinfo {author} {\bibfnamefont {A.}~\bibnamefont {Szolnoki}}\ and\ \bibinfo {author} {\bibfnamefont {M.}~\bibnamefont {Perc}},\ }\bibfield  {title} {\bibinfo {title} {Conditional strategies and the evolution of cooperation in spatial public goods games},\ }\href {https://doi.org/10.1103/physreve.85.026104} {\bibfield  {journal} {\bibinfo  {journal} {Physical Review E}\ }\textbf {\bibinfo {volume} {85}},\ \bibinfo {pages} {026104} (\bibinfo {year} {2012})}\BibitemShut {NoStop}%
\bibitem [{\citenamefont {Chaudhuri}(2011)}]{chaudhuri_sustaining_2011}%
  \BibitemOpen
  \bibfield  {author} {\bibinfo {author} {\bibfnamefont {A.}~\bibnamefont {Chaudhuri}},\ }\bibfield  {title} {\bibinfo {title} {Sustaining cooperation in laboratory public goods experiments: a selective survey of the literature},\ }\href {https://doi.org/10.1007/s10683-010-9257-1} {\bibfield  {journal} {\bibinfo  {journal} {Experimental Economics}\ }\textbf {\bibinfo {volume} {14}},\ \bibinfo {pages} {47} (\bibinfo {year} {2011})}\BibitemShut {NoStop}%
\bibitem [{\citenamefont {Battu}\ and\ \citenamefont {Srinivasan}(2020)}]{Balaraju2020}%
  \BibitemOpen
  \bibfield  {author} {\bibinfo {author} {\bibfnamefont {B.}~\bibnamefont {Battu}}\ and\ \bibinfo {author} {\bibfnamefont {N.}~\bibnamefont {Srinivasan}},\ }\bibfield  {title} {\bibinfo {title} {Evolution of conditional cooperation in public good games},\ }\href {https://doi.org/10.1098/rsos.191567} {\bibfield  {journal} {\bibinfo  {journal} {Royal Society Open Science}\ }\textbf {\bibinfo {volume} {7}},\ \bibinfo {pages} {191567} (\bibinfo {year} {2020})}\BibitemShut {NoStop}%
\bibitem [{\citenamefont {Richerson}\ and\ \citenamefont {Boyd}(2006)}]{boyd_richerson_2006}%
  \BibitemOpen
  \bibfield  {author} {\bibinfo {author} {\bibfnamefont {P.~J.}\ \bibnamefont {Richerson}}\ and\ \bibinfo {author} {\bibfnamefont {R.}~\bibnamefont {Boyd}},\ }\href {http://www.amazon.com/exec/obidos/redirect?tag=citeulike07-20&path=ASIN/0226712125} {\emph {\bibinfo {title} {Not by Genes Alone: How Culture Transformed Human Evolution}}}\ (\bibinfo  {publisher} {University Of Chicago Press},\ \bibinfo {year} {2006})\BibitemShut {NoStop}%
\bibitem [{\citenamefont {Mesoudi}(2009)}]{mesoudi_how_2009}%
  \BibitemOpen
  \bibfield  {author} {\bibinfo {author} {\bibfnamefont {A.}~\bibnamefont {Mesoudi}},\ }\bibfield  {title} {\bibinfo {title} {How cultural evolutionary theory can inform social psychology and vice versa.},\ }\href {https://doi.org/10.1037/a0017062} {\bibfield  {journal} {\bibinfo  {journal} {Psychological Review}\ }\textbf {\bibinfo {volume} {116}},\ \bibinfo {pages} {929} (\bibinfo {year} {2009})},\ \bibinfo {note} {place: US Publisher: American Psychological Association}\BibitemShut {NoStop}%
\bibitem [{\citenamefont {Harsanyi}(1967)}]{harsanyi_games_1967}%
  \BibitemOpen
  \bibfield  {author} {\bibinfo {author} {\bibfnamefont {J.~C.}\ \bibnamefont {Harsanyi}},\ }\bibfield  {title} {\bibinfo {title} {Games with {Incomplete} {Information} {Played} by “{Bayesian}” {Players}, {I}–{III} {Part} {I}. {The} {Basic} {Model}},\ }\href {https://doi.org/10.1287/mnsc.14.3.159} {\bibfield  {journal} {\bibinfo  {journal} {Management Science}\ }\textbf {\bibinfo {volume} {14}},\ \bibinfo {pages} {159} (\bibinfo {year} {1967})},\ \bibinfo {note} {publisher: INFORMS}\BibitemShut {NoStop}%
\bibitem [{\citenamefont {Aumann}\ \emph {et~al.}(1995)\citenamefont {Aumann}, \citenamefont {Maschler},\ and\ \citenamefont {Stearns}}]{aumann_repeated_1995}%
  \BibitemOpen
  \bibfield  {author} {\bibinfo {author} {\bibfnamefont {R.}~\bibnamefont {Aumann}}, \bibinfo {author} {\bibfnamefont {M.}~\bibnamefont {Maschler}},\ and\ \bibinfo {author} {\bibfnamefont {R.}~\bibnamefont {Stearns}},\ }\href {https://books.google.es/books?id=xaa7xZ-WGBsC} {\emph {\bibinfo {title} {Repeated {Games} with {Incomplete} {Information}}}}\ (\bibinfo  {publisher} {MIT Press},\ \bibinfo {year} {1995})\BibitemShut {NoStop}%
\bibitem [{\citenamefont {Forges}(2012)}]{hal-02447604}%
  \BibitemOpen
  \bibfield  {author} {\bibinfo {author} {\bibfnamefont {F.}~\bibnamefont {Forges}},\ }\bibfield  {title} {\bibinfo {title} {{Folk theorems for Bayesian (public good) games}},\ }in\ \href {https://hal.science/hal-02447604} {\emph {\bibinfo {booktitle} {{Thirteenth annual conference (PET12) of the Association for Public Economic Theory (APET)}}}}\ (\bibinfo {address} {Taipei, China},\ \bibinfo {year} {2012})\ p.~\bibinfo {pages} {30}\BibitemShut {NoStop}%
\bibitem [{\citenamefont {Achtziger}\ and\ \citenamefont {Alós-Ferrer}(2014)}]{achtziger_fast_2014}%
  \BibitemOpen
  \bibfield  {author} {\bibinfo {author} {\bibfnamefont {A.}~\bibnamefont {Achtziger}}\ and\ \bibinfo {author} {\bibfnamefont {C.}~\bibnamefont {Alós-Ferrer}},\ }\bibfield  {title} {\bibinfo {title} {Fast or {Rational}? {A} {Response}-{Times} {Study} of {Bayesian} {Updating}},\ }\href {https://doi.org/10.1287/mnsc.2013.1793} {\bibfield  {journal} {\bibinfo  {journal} {Management Science}\ }\textbf {\bibinfo {volume} {60}},\ \bibinfo {pages} {923} (\bibinfo {year} {2014})}\BibitemShut {NoStop}%
\bibitem [{\citenamefont {El-Gamal}\ \emph {et~al.}(1995)\citenamefont {El-Gamal}, \citenamefont {Grether},\ and\ \citenamefont {Grether}}]{ArePeopleBayesian}%
  \BibitemOpen
  \bibfield  {author} {\bibinfo {author} {\bibfnamefont {M.~A.}\ \bibnamefont {El-Gamal}}, \bibinfo {author} {\bibfnamefont {D.~M.}\ \bibnamefont {Grether}},\ and\ \bibinfo {author} {\bibfnamefont {D.~M.}\ \bibnamefont {Grether}},\ }\bibfield  {title} {\bibinfo {title} {Are {People} {Bayesian}? {Uncovering} {Behavioral} {Strategies}},\ }\href {https://doi.org/10.1080/01621459.1995.10476620} {\bibfield  {journal} {\bibinfo  {journal} {Journal of the American Statistical Association}\ }\textbf {\bibinfo {volume} {90}},\ \bibinfo {pages} {1137} (\bibinfo {year} {1995})}\BibitemShut {NoStop}%
\bibitem [{\citenamefont {Sánchez}(2018)}]{Sanchez_2018}%
  \BibitemOpen
  \bibfield  {author} {\bibinfo {author} {\bibfnamefont {A.}~\bibnamefont {Sánchez}},\ }\bibfield  {title} {\bibinfo {title} {Physics of human cooperation: experimental evidence and theoretical models},\ }\href {https://doi.org/10.1088/1742-5468/aaa388} {\bibfield  {journal} {\bibinfo  {journal} {Journal of Statistical Mechanics: Theory and Experiment}\ }\textbf {\bibinfo {volume} {2018}},\ \bibinfo {pages} {024001} (\bibinfo {year} {2018})}\BibitemShut {NoStop}%
\bibitem [{\citenamefont {Dunbar}(1992)}]{DUNBAR1992469}%
  \BibitemOpen
  \bibfield  {author} {\bibinfo {author} {\bibfnamefont {R.}~\bibnamefont {Dunbar}},\ }\bibfield  {title} {\bibinfo {title} {Neocortex size as a constraint on group size in primates},\ }\href {https://doi.org/https://doi.org/10.1016/0047-2484(92)90081-J} {\bibfield  {journal} {\bibinfo  {journal} {Journal of Human Evolution}\ }\textbf {\bibinfo {volume} {22}},\ \bibinfo {pages} {469} (\bibinfo {year} {1992})}\BibitemShut {NoStop}%
\bibitem [{\citenamefont {Choi}(1997)}]{Choi97}%
  \BibitemOpen
  \bibfield  {author} {\bibinfo {author} {\bibfnamefont {J.~P.}\ \bibnamefont {Choi}},\ }\bibfield  {title} {\bibinfo {title} {Herd behavior, the "penguin effect," and the suppression of informational diffusion: An analysis of informational externalities and payoff interdependency},\ }\href@noop {} {\bibfield  {journal} {\bibinfo  {journal} {The RAND Journal of Economics}\ }\textbf {\bibinfo {volume} {28}},\ \bibinfo {pages} {407} (\bibinfo {year} {1997})}\BibitemShut {NoStop}%
\bibitem [{\citenamefont {Diamond}\ and\ \citenamefont {Dybvig}(1983)}]{Diamond83}%
  \BibitemOpen
  \bibfield  {author} {\bibinfo {author} {\bibfnamefont {D.~W.}\ \bibnamefont {Diamond}}\ and\ \bibinfo {author} {\bibfnamefont {P.~H.}\ \bibnamefont {Dybvig}},\ }\bibfield  {title} {\bibinfo {title} {Bank runs, deposit insurance, and liquidity},\ }\href {https://doi.org/10.1086/261155} {\bibfield  {journal} {\bibinfo  {journal} {Journal of Political Economy}\ }\textbf {\bibinfo {volume} {91}},\ \bibinfo {pages} {401} (\bibinfo {year} {1983})}\BibitemShut {NoStop}%
\bibitem [{\citenamefont {Scharfstein}\ and\ \citenamefont {Stein}(1990)}]{Scharfstein90}%
  \BibitemOpen
  \bibfield  {author} {\bibinfo {author} {\bibfnamefont {D.~S.}\ \bibnamefont {Scharfstein}}\ and\ \bibinfo {author} {\bibfnamefont {J.~C.}\ \bibnamefont {Stein}},\ }\bibfield  {title} {\bibinfo {title} {Herd behavior and investment},\ }\href@noop {} {\bibfield  {journal} {\bibinfo  {journal} {Amercian Economic Review}\ }\textbf {\bibinfo {volume} {80}},\ \bibinfo {pages} {465} (\bibinfo {year} {1990})}\BibitemShut {NoStop}%
\bibitem [{\citenamefont {Manski}(2000)}]{Manski2000}%
  \BibitemOpen
  \bibfield  {author} {\bibinfo {author} {\bibfnamefont {C.~F.}\ \bibnamefont {Manski}},\ }\bibfield  {title} {\bibinfo {title} {Economic analysis of social interactions},\ }\href {https://doi.org/10.1257/jep.14.3.115} {\bibfield  {journal} {\bibinfo  {journal} {Journal of Economic Perspectives}\ }\textbf {\bibinfo {volume} {14}},\ \bibinfo {pages} {115–136} (\bibinfo {year} {2000})}\BibitemShut {NoStop}%
\bibitem [{\citenamefont {Banerjee}(1992)}]{Banerjee92}%
  \BibitemOpen
  \bibfield  {author} {\bibinfo {author} {\bibfnamefont {A.~V.}\ \bibnamefont {Banerjee}},\ }\bibfield  {title} {\bibinfo {title} {A simple model of herd behavior},\ }\href {http://www.jstor.org/stable/2118364} {\bibfield  {journal} {\bibinfo  {journal} {The Quarterly Journal of Economics}\ }\textbf {\bibinfo {volume} {107}},\ \bibinfo {pages} {797} (\bibinfo {year} {1992})}\BibitemShut {NoStop}%
\bibitem [{\citenamefont {Bikhchandani}\ \emph {et~al.}(1992)\citenamefont {Bikhchandani}, \citenamefont {Hirshleifer},\ and\ \citenamefont {Welch}}]{Bikhchandani92}%
  \BibitemOpen
  \bibfield  {author} {\bibinfo {author} {\bibfnamefont {S.}~\bibnamefont {Bikhchandani}}, \bibinfo {author} {\bibfnamefont {D.}~\bibnamefont {Hirshleifer}},\ and\ \bibinfo {author} {\bibfnamefont {I.}~\bibnamefont {Welch}},\ }\bibfield  {title} {\bibinfo {title} {A theory of fads, fashion, custom, and cultural change as informational cascades},\ }\href {http://www.jstor.org/stable/2138632} {\bibfield  {journal} {\bibinfo  {journal} {Journal of Political Economy}\ }\textbf {\bibinfo {volume} {100}},\ \bibinfo {pages} {992} (\bibinfo {year} {1992})}\BibitemShut {NoStop}%
\bibitem [{\citenamefont {Welch}(1992)}]{Welch92}%
  \BibitemOpen
  \bibfield  {author} {\bibinfo {author} {\bibfnamefont {I.}~\bibnamefont {Welch}},\ }\bibfield  {title} {\bibinfo {title} {Sequential sales, learning, and cascades},\ }\href {https://doi.org/https://doi.org/10.1111/j.1540-6261.1992.tb04406.x} {\bibfield  {journal} {\bibinfo  {journal} {The Journal of Finance}\ }\textbf {\bibinfo {volume} {47}},\ \bibinfo {pages} {695} (\bibinfo {year} {1992})},\ \Eprint {https://arxiv.org/abs/https://onlinelibrary.wiley.com/doi/pdf/10.1111/j.1540-6261.1992.tb04406.x} {https://onlinelibrary.wiley.com/doi/pdf/10.1111/j.1540-6261.1992.tb04406.x} \BibitemShut {NoStop}%
\bibitem [{\citenamefont {Corazzini}\ and\ \citenamefont {Greiner}(2007)}]{CORAZZINI200774}%
  \BibitemOpen
  \bibfield  {author} {\bibinfo {author} {\bibfnamefont {L.}~\bibnamefont {Corazzini}}\ and\ \bibinfo {author} {\bibfnamefont {B.}~\bibnamefont {Greiner}},\ }\bibfield  {title} {\bibinfo {title} {Herding, social preferences and (non-)conformity},\ }\href {https://doi.org/https://doi.org/10.1016/j.econlet.2007.02.024} {\bibfield  {journal} {\bibinfo  {journal} {Economics Letters}\ }\textbf {\bibinfo {volume} {97}},\ \bibinfo {pages} {74} (\bibinfo {year} {2007})}\BibitemShut {NoStop}%
\bibitem [{\citenamefont {Goeree}\ and\ \citenamefont {Yariv}(2015)}]{Goeree2015}%
  \BibitemOpen
  \bibfield  {author} {\bibinfo {author} {\bibfnamefont {J.~K.}\ \bibnamefont {Goeree}}\ and\ \bibinfo {author} {\bibfnamefont {L.}~\bibnamefont {Yariv}},\ }\bibfield  {title} {\bibinfo {title} {Conformity in the lab},\ }\href {https://doi.org/10.1007/s40881-015-0001-7} {\bibfield  {journal} {\bibinfo  {journal} {Journal of the Economic Science Association}\ }\textbf {\bibinfo {volume} {1}},\ \bibinfo {pages} {15} (\bibinfo {year} {2015})}\BibitemShut {NoStop}%
\bibitem [{\citenamefont {Jones}(1984)}]{Jones94}%
  \BibitemOpen
  \bibfield  {author} {\bibinfo {author} {\bibfnamefont {S.~R.}\ \bibnamefont {Jones}},\ }\href@noop {} {\emph {\bibinfo {title} {The economics of conformism}}}\ (\bibinfo  {publisher} {Blackwell, Oxford},\ \bibinfo {year} {1984})\BibitemShut {NoStop}%
\bibitem [{\citenamefont {Bernheim}(1994)}]{Bernheim94}%
  \BibitemOpen
  \bibfield  {author} {\bibinfo {author} {\bibfnamefont {B.~D.}\ \bibnamefont {Bernheim}},\ }\bibfield  {title} {\bibinfo {title} {A theory of conformity},\ }\href {https://EconPapers.repec.org/RePEc:ucp:jpolec:v:102:y:1994:i:5:p:841-77} {\bibfield  {journal} {\bibinfo  {journal} {Journal of Political Economy}\ }\textbf {\bibinfo {volume} {102}},\ \bibinfo {pages} {841} (\bibinfo {year} {1994})}\BibitemShut {NoStop}%
\end{thebibliography}%

\end{document}